\documentclass[twoside,twocolumn,9pt]{article}
\usepackage{extsizes}
\usepackage[super,sort&compress,comma]{natbib} 
\usepackage[version=3]{mhchem}
\usepackage[left=1.5cm, right=1.5cm, top=1.785cm, bottom=2.0cm]{geometry}
\usepackage{balance}
\usepackage{mathptmx}
\usepackage{sectsty}
\usepackage{graphicx} 
\usepackage{lastpage}
\usepackage[format=plain,justification=justified,singlelinecheck=false,font={stretch=1.125,small,sf},labelfont=bf,labelsep=space]{caption}
\usepackage{float}
\usepackage{fancyhdr}
\usepackage{fnpos}
\usepackage[english]{babel}
\addto{\captionsenglish}{%
  
}
\usepackage{array}
\usepackage{droidsans}
\usepackage{charter}
\usepackage[T1]{fontenc}
\usepackage[usenames,dvipsnames]{xcolor}
\usepackage{setspace}
\usepackage[compact]{titlesec}
\usepackage{hyperref}

\usepackage{comment}
\usepackage{epstopdf}

\definecolor{cream}{RGB}{222,217,201}

\usepackage{afterpage}
\usepackage{longtable}
\usepackage{lipsum, babel}

\begin{document}

\pagestyle{fancy}
\thispagestyle{plain}
\fancypagestyle{plain}{
\renewcommand{\headrulewidth}{0pt}
}

\makeFNbottom
\makeatletter
\renewcommand\LARGE{\@setfontsize\LARGE{15pt}{17}}
\renewcommand\Large{\@setfontsize\Large{12pt}{14}}
\renewcommand\large{\@setfontsize\large{10pt}{12}}
\renewcommand\footnotesize{\@setfontsize\footnotesize{7pt}{10}}
\makeatother

\renewcommand{\thefootnote}{\fnsymbol{footnote}}
\renewcommand\footnoterule{\vspace*{1pt}%
\color{cream}\hrule width 3.5in height 0.4pt \color{black}\vspace*{5pt}} 
\setcounter{secnumdepth}{5}

\makeatletter 
\renewcommand\@biblabel[1]{#1}            
\renewcommand\@makefntext[1]%
{\noindent\makebox[0pt][r]{\@thefnmark\,}#1}
\makeatother 
\renewcommand{\figurename}{\small{Fig.}~}
\sectionfont{\sffamily\Large}
\subsectionfont{\normalsize}
\subsubsectionfont{\bf}
\setstretch{1.125} 
\setlength{\skip\footins}{0.8cm}
\setlength{\footnotesep}{0.25cm}
\setlength{\jot}{10pt}
\titlespacing*{\section}{0pt}{4pt}{4pt}
\titlespacing*{\subsection}{0pt}{15pt}{1pt}

\fancyfoot{}
\fancyfoot[LO,RE]{\vspace{-7.1pt}\includegraphics[height=9pt]{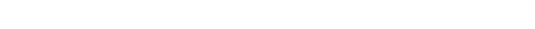}}
\fancyfoot[CO]{\vspace{-7.1pt}\hspace{13.2cm}\includegraphics{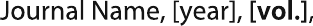}}
\fancyfoot[CE]{\vspace{-7.2pt}\hspace{-14.2cm}\includegraphics{head_foot/RF}}
\fancyfoot[RO]{\footnotesize{\sffamily{1--\pageref{LastPage} ~\textbar  \hspace{2pt}\thepage}}}
\fancyfoot[LE]{\footnotesize{\sffamily{\thepage~\textbar\hspace{3.45cm} 1--\pageref{LastPage}}}}
\fancyhead{}
\renewcommand{\headrulewidth}{0pt} 
\renewcommand{\footrulewidth}{0pt}
\setlength{\arrayrulewidth}{1pt}
\setlength{\columnsep}{6.5mm}
\setlength\bibsep{1pt}

\makeatletter 
\newlength{\figrulesep} 
\setlength{\figrulesep}{0.5\textfloatsep} 

\newcommand{\topfigrule}{\vspace*{-1pt}%
\noindent{\color{cream}\rule[-\figrulesep]{\columnwidth}{1.5pt}} }

\newcommand{\botfigrule}{\vspace*{-2pt}%
\noindent{\color{cream}\rule[\figrulesep]{\columnwidth}{1.5pt}} }

\newcommand{\dblfigrule}{\vspace*{-1pt}%
\noindent{\color{cream}\rule[-\figrulesep]{\textwidth}{1.5pt}} }

\makeatother

\twocolumn[
  \begin{@twocolumnfalse}
{\includegraphics[height=30pt]{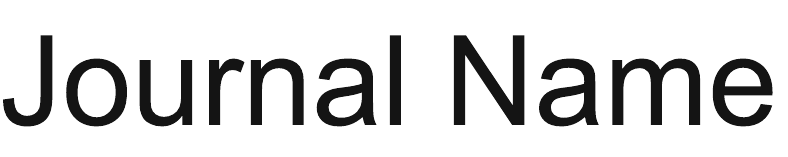}\hfill\raisebox{0pt}[0pt][0pt]{\includegraphics[height=55pt]{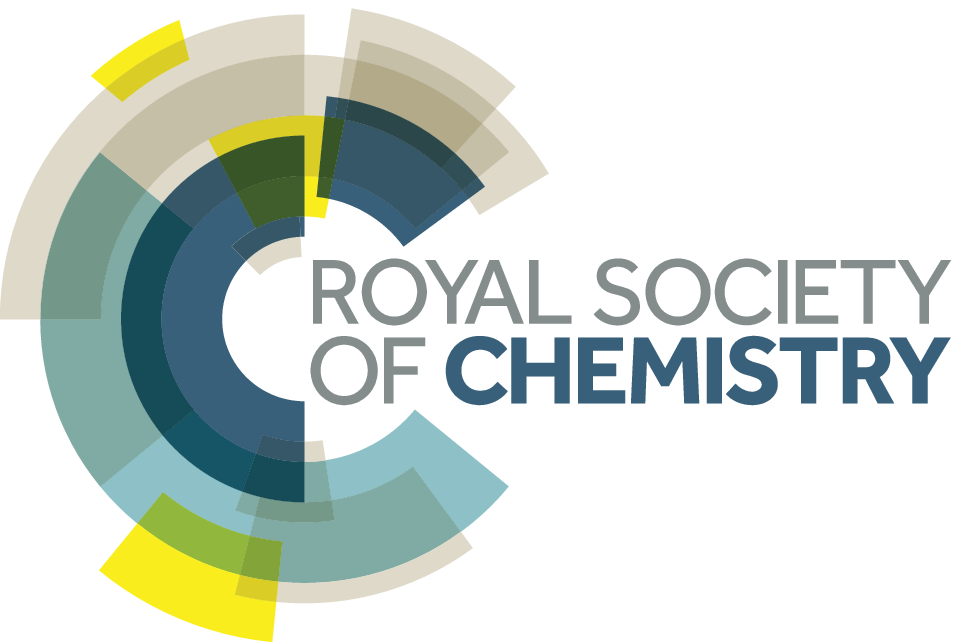}}\\[1ex]
\includegraphics[width=18.5cm]{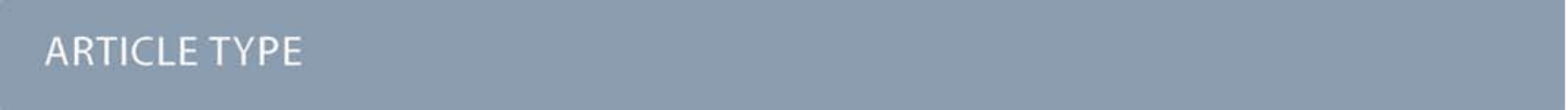}}\par
\vspace{1em}
\sffamily
\begin{tabular}{m{4.5cm} p{13.5cm} }

\includegraphics{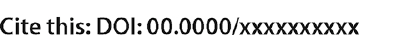} & \noindent\LARGE{\textbf{Insights on the coupling between vibronically active molecular vibrations and lattice phonons in molecular nanomagnets}} \\
\vspace{0.3cm} & \vspace{0.3cm} \\

 & \noindent\large{Aman Ullah,\textit{$^{a}$} Jos\'e J. Baldov\'i,\textit{$^{a}$} Alejandro Gaita-Ari\~no$^{\ast}$\textit{$^{a}$} and Eugenio Coronado\textit{$^{a}$}} \\

\includegraphics{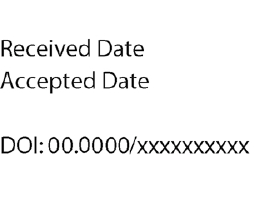} & \noindent\normalsize{Spin-lattice relaxation is a key open problem to understand the spin dynamics of single-molecule magnets and molecular spin qubits. While modelling the coupling between spin states and local vibrations allows to determine the more relevant molecular vibrations for spin relaxation, this is not sufficient to explain how energy is dissipated towards the thermal bath. Herein, we employ a simple and efficient model to examine the coupling of local vibrational modes with long-wavelength longitudinal and transverse phonons in the clock-like spin qubit [Ho(W$_5$O$_{18}$)$_2$]$^{9-}$. We find that in crystals of this polyoxometalate the vibrational mode previously found to be vibronically active at low temperature does not couple significantly to lattice phonons. This means that further intramolecular energy transfer via anharmonic vibrations is necessary for spin relaxation in this system. Finally, we discuss implications for the spin-phonon coupling of [Ho(W$_5$O$_{18}$)$_2$]$^{9-}$ deposited on a MgO (001) substrate, offering a simple methodology that can be extrapolated to estimate the effects on spin relaxation of different surfaces, including 2D materials.} \\

\end{tabular}

 \end{@twocolumnfalse} \vspace{0.6cm}

  ]

\renewcommand*\rmdefault{bch}\normalfont\upshape
\rmfamily
\section*{}
\vspace{-1cm}


\footnotetext{\textit{$^{a}$~Instituto de Ciencia Molecular, Universitat de València, Catedrático José Beltrán Martínez, 2, Paterna 46980, Spain; E-mail: gaita@uv.es}}

\footnotetext{\dag~Electronic Supplementary Information (ESI) available: [Supplementary Tables including vibronic and vibronic-phonon coupling for 110 normal modes]. See DOI: 00.0000/00000000.}




\section*{Introduction}

Molecular magnetism is a field that has seen remarkable progress over the last three decades. This has resulted in a rich bibliography that covers hundreds of coordination complexes and requires specialized data science tools to properly visualize the experimental results.\cite{Aromi2019,Long2018,coronado2020molecular,SIMDAVIS} Starting from single-molecule magnets (SMMs) based on transition metal clusters,\cite{Sessoli1993} the discovery of the so-called single-ion magnets (SIMs), mainly based on lanthanides, allowed the obtention of improved magnetic properties, which stem from the intrinsically large magnetic moment and spin-orbit coupling of rare earth ions.\cite{ishikawa2003lanthanide,jiang2011} Eventually, the SIM strategy lead to the design of coordination spheres resulting in record potential energy barriers and hysteresis temperatures above liquid nitrogen.\cite{Harriman2017,Guo2018,Li2019} This was possible, at least in part, because this path provides a more intuitive picture to model and rationalize the magnetic anisotropy of a single magnetic ion and its coordination environment.\cite{Rinehart2011,Ungur2016,Mcadams2017} While new avenues are still being explored to design and achieve higher potential energy barriers, most theoretical research now focuses on decoupling spin states from molecular vibrations using first principles calculations.\cite{Lunghi2017,ullah2019silico} Indeed, the step that allowed rising the hysteresis temperature from 60 K to 80 K involved precisely a judicious modification of the ligands, not to rise the barrier but to eliminate a vibronically active vibrational mode, i.e. a mode that coupled significantly to spin states.\cite{Goodwin2017hysteresis,McClain2018,Guo2018} It has been shown in different ways that spin states in magnetic molecules couple most strongly with local vibrations,\cite{Atzori2016,Goodwin2017hysteresis,Rosaleny2019} rather than with lattice phonons as is the case for other magnetic solids.\cite{ONeal2019} Moreover, the details of the spin relaxation mechanism in systems with the highest hysteresis temperatures are being actively investigated, in terms of whether Orbach or Raman are the dominant processes.\cite{Giansiracusa2019,Castro2020} However, there is still an open question in determining how local vibrations couple with lattice phonons to allow for the energy to flow from the spins of the molecule to the thermal bath of the solid. More generally, understanding the coupling between spin states, molecular vibrations and lattice phonons will also help to understand the dynamics of molecular ferroelectrics\cite{Long2020} and hybrid molecular solid-state materials. 

A recent work combining 4D Inelastic Neutron Scattering (4D-INS) with DFT calculations has demonstrated that explicit analysis of phonon eigenvectors is necessary to properly estimate spin-phonon coupling. This involves not only localized molecular vibrations, related to optical phonon, but also extended lattice vibrations, also known as acoustic phonons.\cite{garlatti2020unveiling} The participation of low-lying optical phonons, i.e. molecular vibrations, in spin relaxation is more obvious, since the distortions couple strongly with the spin Hamiltonian. Less obvious, but also crucial, is the consideration of anticrossings between low-lying optical branches and acoustic phonons. These can be experimentally detected only by measuring phonon dispersions and it is well-known that can cause a decrease on the phonon lifetimes\cite{Lee2006,Christensen2008,Toberer2011}. The mixing of optical and acoustic phonon eigenvectors due to these anticrossings results in an effective communication of the spin states with the thermal bath, thus enhancing magnetic relaxation in molecular nanomagnets and reducing T$_1$ in molecular spin qubits.\cite{Lunghi2017CS}

In order to advance towards a solution, it seems adequate to start from a model molecular nanomagnet that has been extensively studied in the past few years. For that, we focus our attention on [Ho(W$_5$O$_{18}$)$_2$]$^{9-}$, a polyoxometalate system that displays so-called Atomic Clock Transitions, which protect qubit states from magnetic noise. This allows for coherent operation at high concentrations.\cite{Shiddiq2016} Over the years, the spin energy level structure of this system has been widely characterized, both from the experimental and theoretical points of view.\cite{Shiozaki1996,AlDamen2009,Ghosh2012,Vonci2017, escalera2019unveiling} In particular, recent studies have been able to determine the couplings between spin states and molecular vibrations or distortions. In particular, a theoretical characterization at CASSCF level has been employed to rationalize experiments of quantum coherent electrical control of spins\cite{Liu2021} and magneto-infrared spectroscopy.\cite{Blockmon2021} In Liu et al.,\cite{Liu2021} the molecular structure was distorted by applying an external electric field, and the coupling of this distortion with the Crystal Field Hamiltonian was modeled by expanding it in the basis of normal modes of vibration. A first principles model was able to reproduce the experimentally determined effect of the electric field on the tunneling gap. In Blockmon et al.,\cite{Blockmon2021} modelling of vibronic coupling allowed us to reproduce the observed dependence of the infrared spectra at rising magnetic fields. Starting from these successful schemes for the vibronic coupling in Na$_{9}$[Ho(W$_5$O$_{18}$)$_2$]$\cdot$35H$_2$O, herein we develop a theoretical model to explore the missing link between local vibrations and the thermal bath, namely vibronic-phonon coupling.

\section*{Results and discussion}

\subsection*{Modelling vibronic-phonon coupling}

\begin{figure}[tbh!]
\centering
\includegraphics[width=1.0\columnwidth]{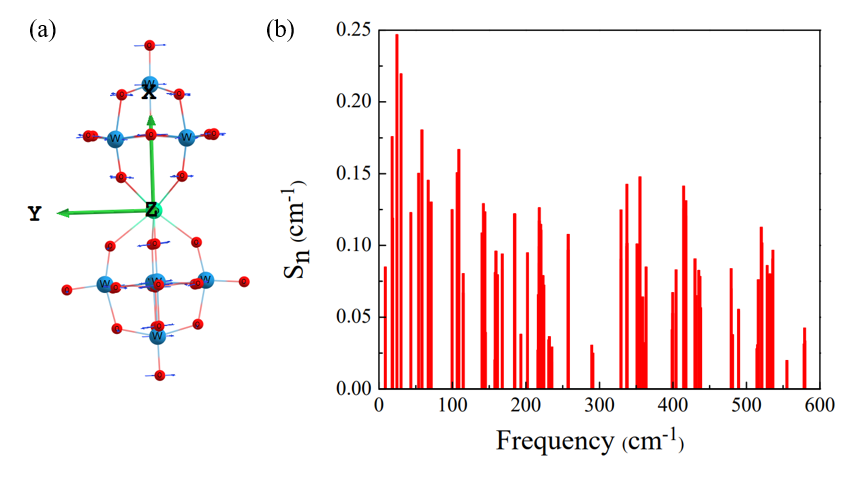}\\
\caption{(a) Structure of the molecular anion [Ho(W$_5$O$_{18}$)$_2$]$^{9-}$, superimposed with vectors indicating the atomic displacements for normal mode $n=1$ and the ($x,y,z$) reference frame. (b) Vibronic coupling $S_n$ vs vibrational frequency for molecular normal modes in [Ho(W$_5$O$_{18}$)$_2$]$^{9-}$.}
\label{HoW10couplings}
\end{figure}

Previous analyses in [Ho(W$_5$O$_{18}$)$_2$]$^{9-}$ have identified the more crucial vibronically active vibrations.\cite{Blockmon2021} Specific $M_J$ spin energy levels couple to various HoO$_8$ rocking and stretching modes. This includes asymmetric HoO$_8$ stretching with cage tilting, and, in particular, the vibrational mode $n=1$, presenting the lowest vibrational frequency $\nu=8.6$~cm$^{-1}$ (Fig.~\ref{HoW10couplings}). This mode can be approximated as a torsion along the near-$C_4$ molecular axis, where the upper moiety rotates clockwise while the lower one rotates anti-clockwise. At the low temperature range, relevant for the clock transition experiments, this vibration is involved in the thermal dependence of the spin-lattice relaxation rate $T_1$.\cite{Blockmon2021} The vibronic coupling $S_n$ for a dense spectrum of vibronically active vibrations, as calculated in that work, is represented in Fig.~\ref{HoW10couplings}. The key question is therefore whether all these vibrational modes, and in particular $n=1$, couple directly to phonons.

A conceptual cornerstone of the model employed in the present work is the fact that we deal with phonons of long-wavelength, which we assume to (i) not depend on the molecular details and (ii) couple only weakly to spin states. To model this kind of phonons one can employ the same approach described previously in the context of strong and weak interacting two level systems in disordered solids.\cite{Gaita2011} This is based on the generally accepted idea that it is often a convenient approximation to distinguish between “longitudinal” and “transverse” acoustic phonons, although in practice they are not pure.\cite{garlatti2020unveiling}

\begin{figure}[tbh!]
\centering
\includegraphics[width=0.63\columnwidth]{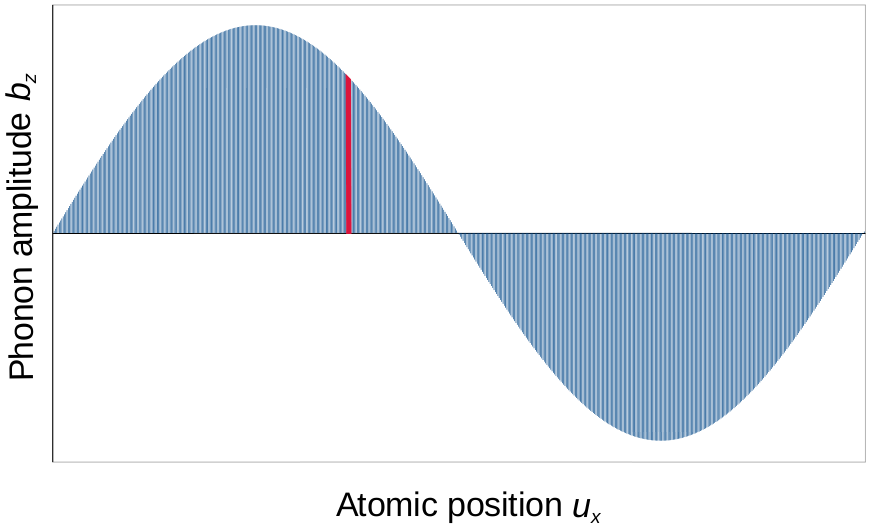}
\includegraphics[width=0.32\columnwidth]{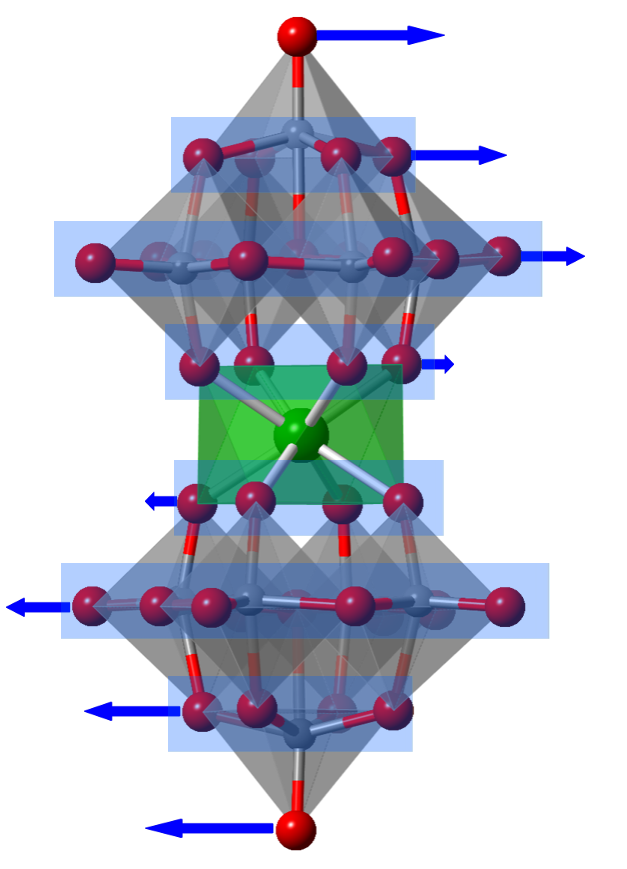}\\
\caption{Left: Long-wavelength transverse phonon $\mu_{zx}$, with phonon amplitude along $z$ and propagating along $x$ (phonon amplitude greatly exagerated to facilitate visualization). Assuming the molecule fits in the width of the red line, each atom suffers a strain $b_z$ along the phonon amplitude direction $z$ that is directly proportional to its position $u_x$ along the propagation direction $x$. Right: Relative effect of transverse phonon $\mu_{zx}$ on [Ho(W$_5$O$_{18}$)$_2$]$^{9-}$ with respect to the central Ho$^{3+}$ ion. Color code: Ho (green), W (gray) and O (red).}
\label{HoW10vectors}
\end{figure}

Let us say we want to model the effect on a local excitation of a phonon field $\mu_{\alpha\beta}$, where $\alpha=x$ is the unit vector of the phonon amplitude and $\beta=x$ is the unit vector of the phonon propagation direction. For simplicity we can take as an example the longitudinal phonon $\mu_{xx}$ along $x$. Assuming that the phonon has a long-wavelength compared with our local excitation, we can just employ a homogeneous lattice contraction by a fraction $b$ along crystallographic direction $x$ to mimic the effect of a longitudinal phonon along the same coordinate. The distorted molecular coordinates $d_\alpha$ can be defined by considering, for each atom, the equilibrium coordinates $e_\alpha$ and the distortion $b$, which is proportional to the position $e_\beta$ along the propagation vector $\beta$: 
\begin{equation}
    d_\alpha = e_\alpha + b\cdot e_\beta
\end{equation}
The application of a longitudinal phonon $\mu_{xx}$ to an atom with equilibrium coordinates $(e_x,e_y,e_z)$ results in the distorted coordinates $(e_x+b\cdot e_x,e_y,e_z)$. As an example for transverse phonon let us take $\mu_{xz}$, which results in the distorted coordinates $(e_x+b\cdot e_z,e_y,e_z)$ (see Fig. \ref{HoW10vectors}). Three longitudinal phonons  $\mu_{xx},\mu_{yy},\mu_{zz}$ and six transverse phonons $\mu_{xy},\mu_{xz},\mu_{yx},\mu_{yz},\mu_{zx},\mu_{zy}$ can be constructed in this way. Locally, each of the three longitudinal phonons acts like a uniaxial strain, while each of the six transverse ones acts as a shear strain. This permits an exploration of the long-wavelength part of the phonon spectrum employing just 9 model phonons. Figure 2 illustrates the relation between phonon amplitude and atomic position for the example of a transverse phonon $\mu_{zx}$, as well as the qualitative effect of such a phonon on the molecular anion [Ho(W$_5$O$_{18}$)$_2$]$^{9-}$, where we take as a criterion that the long axis of the molecule defines the $x$ axis.

To facilitate comparisons one can impose a normalizing condition to the phonon displacements so that the product $\vec{\mu}_{\alpha\beta}\cdot\vec{\mu}_{\alpha\beta}=1$ for any $\alpha$,$\beta$. The vibronic-phonon coupling $S(\mu_{n,\alpha\beta})$ is simply calculated as the scalar product between the normalized displacement vectors defining each normal mode $\vec{n}$ and each of the idealized phonons $\vec{\mu}_{\alpha\beta}$, in absolute value:
\begin{equation}
    S(\mu_{n,\alpha\beta}) = | \vec{n} \cdot \vec{\mu}_{\alpha\beta} |
\end{equation}

\begin{figure*}[tbh!]
\centering
\includegraphics[width=0.85\textwidth]{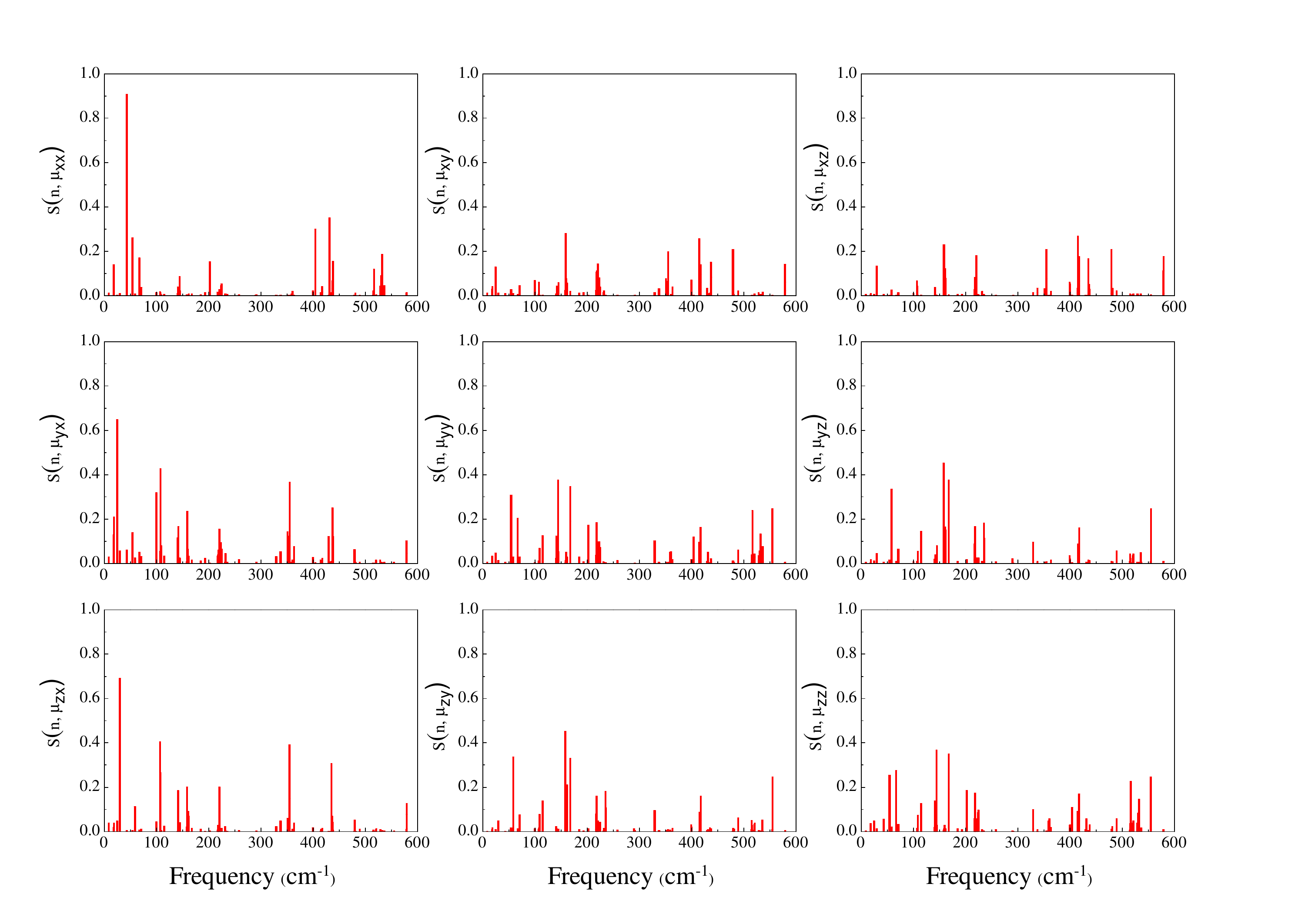}
\caption{Plots of vibronic-phonon coupling vs normal mode energy, overlayed with vibronic coupling of each mode. Rows correspond to the direction of the phonon amplitude of the distortion, columns correspond to the direction of the phonon propagation. The long axis of the molecule defines the $x$ axis. Longitudinal phonons  $\mu_{xx},\mu_{yy},\mu_{zz}$ are thus in the diagonal, and transverse phonons $\mu_{xy},\mu_{xz},\mu_{yx},\mu_{yz},\mu_{zx},\mu_{zy}$ in their corresponding extradiagonal locations.}
\label{vibronic-phonon}
\end{figure*}

In practice, we are estimating the intensity of anticrossings between optical and acoustic phonon branches.  This could be experimentally characterized using 4D-INS as previously shown.\cite{garlatti2020unveiling}

\subsection*{Application to [Ho(W$_5$O$_{18}$)$_2$]$^{9-}$}

 Figure~\ref{vibronic-phonon} shows the coupling of each molecular normal mode of vibration to our 9 model phonons (for full details, see Supplementary Information). Since $S(\mu_{n,\alpha\beta})=1$ is the theoretical maximum for a normal mode that has the exact displacement vector as one of the idealized phonons we defined, it is immediate to see that many vibrations couple significantly to at least a type of phonon, with values $0.1<S(\mu_{n,\alpha\beta})<0.2$. Additionally, vibrations $n=4,5,6$ (24, 30, 43 cm $^{-1}$) couple strongly with the three phonons propagating along the molecular axis $\mu_{xx},\mu_{yx},\mu_{zx}$. At the same time one can notice that the spectrum is much more sparse compared to the one depicting vibronic coupling strength in Fig.~\ref{HoW10couplings}. Note that the values are influenced by the reference frame choice defining the phonons, but that for symmetry reasons all qualitative conclusions are independent of that choice. In particular, vibrations that are orthogonal to all 9 phonons in a given reference frame will consistently be orthogonal for any other reference frame.

 Inspecting the numbers in some detail (see Supplementary Information) we find that for mode $n=1$ all vibronic-phonon couplings $S(\mu_{1,\alpha\beta})$ are of the order of 0.01 or smaller. This means that, even if at low temperature the lowest frequency vibrational mode was found in a previous work to be the one that couples with the spin subsystem and governs the relaxation,\cite{Blockmon2021} it cannot communicate directly with the phonon bath. Thus, an extra step is required to complete the relaxation pathway, probably via the anharmonicity of the molecular vibrations which breaks their exact orthogonality and allows energy transfer between them. 
 
 It is also noticeable that there is a window between 220 and 320 cm$^{-1}$ with practically no coupling between local vibrations and phonons. The molecular design of systems where this window is at low energies would be a novel way of forcing a phonon bottleneck that would facilitate long relaxation times. This idea of minimizing vibronic-phonon coupling for vibrations accessible at low temperature is of course complementary to the well-known strategy of avoiding vibronic coupling for the same energy range.

It would now be possible to map known phonon spectra of relevant surfaces onto our 9 phonon scheme and obtain useful insights. In particular, for illustration purposes let us focus on MgO(001), which was employed to decouple Ho atoms from the underlying Ag substrate in the first single atom magnet experiments.\cite{Natterer2017} Recent studies in this system have already elucidated the role of localized vibrational modes in its spin relaxation.\cite{Donati2020} Calculations on a thin slab model\cite{Shpakov2005} identified surface phonons in the region of 150 cm$^{-1}$, corresponding to out-of plane and in-plane shear vibrations. If we define $x$ as the vector normal to the MgO surface, these out-of plane and in-plane shear vibrations would correspond to phonons of the kind $\mu_{xz}, \mu_{xy}$ and  $\mu_{yz}, \mu_{zy}$ respectively. Our analysis evidences that it would be an advantageous environment for [Ho(W$_5$O$_{18}$)$_2$]$^{9-}$ anions, since the first molecular vibrations which are significantly coupled to this kind of phonons would be $n=5$ at 25-30 cm$^{-1}$. Regarding the orientation of the molecule on the surface, due to the stronger coupling of molecular normal modes with the three phonons propagating along the molecular axis, it would be desirable to orient the molecular axis perpendicular to the surface. A possible approach for this purpose is the covalent functionalization of the molecular nanomagnet using a grafting functional group. Indeed, it has been reported that in
[Ho$\{$Mo$_5$O$_{13}$(OMe)$_4$NNC$_6$H$_4$-$p$-NO$_2\}_2$]$^{3-}$,\cite{baldovi2016single} the molybdenum analogue to [Ho(W$_5$O$_{18}$)$_2$]$^{9-}$, the magnetic ion presents an equivalent coordination sphere in a chemically identical structure, except for the presence of two apical -NO$_2^-$ grafting groups that can be chemically tuned to control the orientation of the molecule in different substrates. Although challenging, alternative chemical strategies such as the use of $\alpha$-cyclodextrins rings, which have been experimentally combined with polyoxopalladates,\cite{stuckart2018host} or rotaxanes that can form surface-attached rotaxanes on gold substrates, \cite{vance2003xas} can be proposed in order to stabilize an off-plane orientation of the molecular axis, which would minimize vibronic-phonon coupling in this molecule, as we have estimated in this work. More generally, this kind of efficient calculation could offer a set of preliminary guidelines in order to reduce the coupling between vibronically active molecular vibrations and surface phonons when the magnetic molecules are deposited onto bidimensional substrates, such as graphene\cite{konstantinov2021} or transition metal dichalcogenides in which charge density waves are also present.\cite{boix2020}

\section*{Conclusions}

Considerable progress has been made in the modelling of spin-lattice relaxation in molecular spin qubits and molecular nanomagnets. However, most of the state-of-the-art models are still essentially zero-dimensional and thus fundamentally unable of describing the energy dissipation to the thermal bath, and only extremelly challenging combinations of state of the art experiments and theory can explicitly unveil these mechanisms.\cite{garlatti2020unveiling} We have employed a model to offer an estimate of the coupling between local vibrational modes and long-wavelength phonons to the clock-like spin qubit Na$_9$HoW$_{10}$O$_{36}\cdot x$H$_2$O. We found that the first vibrational mode, which presents significant vibronic coupling and had been attributed to play a crucial role in spin-lattice relaxation does not couple significantly to long-wavelength lattice phonons, neither longitudinal nor transverse, thus revealing that further intramolecular energy transfer via anharmonic vibrations is involved in the spin relaxation process. As a more general insight, this simple model provides information on the relation of molecular vibrations vis-\`a-vis long-wave phonons. This can be obtained from a very efficient calculation and even in some cases from visual inspection and can be of interest in the design of more robust molecular nanomagnets and spin qubits. This can be especially valuable for the case of single molecules deposited on well-known surfaces, where experiments are challenging and can benefit from theoretical insights.

\section*{Author Contributions}
Conceptualization and methodology: A.G.A.
Investigation, software and formal analysis: A.U.
Supervision: J.J.B. and A.G.A.
Original draft: A.G.A.
Reviewing and editing: E.C. and J.J.B.

\section*{Conflicts of interest}
There are no conflicts to declare.

\section*{Acknowledgements}
We acknowledge funding by the EU (ERC-2014-CoG-647301 DECRESIM, ERC-2018-AdG-788222 MOL-2D, the SUMO QUANTERA Project and FET-OPEN grant 862893 FATMOLS); the Spanish MCIU (grant MAT2017-89993-R and CTQ2017-89528-P cofinanced by FEDER; the Unit of excellence `Mar\'ia de Maeztu' CEX2019-000919-M); the Generalitat Valenciana (PROMETEO/2019/066 and PROMETEO/2017/066, SEJI/2018/035 and grant CDEIGENT/2019/022).

\bibliography{rsc}

\providecommand*{\mcitethebibliography}{\thebibliography}
\csname @ifundefined\endcsname{endmcitethebibliography}
{\let\endmcitethebibliography\endthebibliography}{}
\begin{mcitethebibliography}{45}
\providecommand*{\natexlab}[1]{#1}
\providecommand*{\mciteSetBstSublistMode}[1]{}
\providecommand*{\mciteSetBstMaxWidthForm}[2]{}
\providecommand*{\mciteBstWouldAddEndPuncttrue}
  {\def\EndOfBibitem{\unskip.}}
\providecommand*{\mciteBstWouldAddEndPunctfalse}
  {\let\EndOfBibitem\relax}
\providecommand*{\mciteSetBstMidEndSepPunct}[3]{}
\providecommand*{\mciteSetBstSublistLabelBeginEnd}[3]{}
\providecommand*{\EndOfBibitem}{}
\mciteSetBstSublistMode{f}
\mciteSetBstMaxWidthForm{subitem}
{(\emph{\alph{mcitesubitemcount}})}
\mciteSetBstSublistLabelBeginEnd{\mcitemaxwidthsubitemform\space}
{\relax}{\relax}

\bibitem[Aromí and Roubeau(2019)]{Aromi2019}
G.~Aromí and O.~Roubeau, in \emph{Handbook on the Physics and Chemistry of
  Rare Earths}, Elsevier, 2019, vol.~56, pp. 1--54\relax
\mciteBstWouldAddEndPuncttrue
\mciteSetBstMidEndSepPunct{\mcitedefaultmidpunct}
{\mcitedefaultendpunct}{\mcitedefaultseppunct}\relax
\EndOfBibitem
\bibitem[Long \emph{et~al.}(2018)Long, Guari, Ferreira, Carlos, and
  Larionova]{Long2018}
J.~Long, Y.~Guari, R.~A. Ferreira, L.~D. Carlos and J.~Larionova,
  \emph{Coordination Chemistry Reviews}, 2018, \textbf{363}, 57--70\relax
\mciteBstWouldAddEndPuncttrue
\mciteSetBstMidEndSepPunct{\mcitedefaultmidpunct}
{\mcitedefaultendpunct}{\mcitedefaultseppunct}\relax
\EndOfBibitem
\bibitem[Coronado(2020)]{coronado2020molecular}
E.~Coronado, \emph{Nature Reviews Materials}, 2020, \textbf{5}, 87--104\relax
\mciteBstWouldAddEndPuncttrue
\mciteSetBstMidEndSepPunct{\mcitedefaultmidpunct}
{\mcitedefaultendpunct}{\mcitedefaultseppunct}\relax
\EndOfBibitem
\bibitem[Duan \emph{et~al.}(2021)Duan, Coutinho, Rosaleny, Cardona-Serra,
  Baldoví, and Gaita-Ariño]{SIMDAVIS}
Y.~Duan, J.~T. Coutinho, L.~E. Rosaleny, S.~Cardona-Serra, J.~J. Baldoví and
  A.~Gaita-Ariño, \emph{arXiv}, 2021,  2103.03199\relax
\mciteBstWouldAddEndPuncttrue
\mciteSetBstMidEndSepPunct{\mcitedefaultmidpunct}
{\mcitedefaultendpunct}{\mcitedefaultseppunct}\relax
\EndOfBibitem
\bibitem[Sessoli \emph{et~al.}(1993)Sessoli, Gatteschi, Caneschi, and
  Novak]{Sessoli1993}
R.~Sessoli, D.~Gatteschi, A.~Caneschi and M.~A. Novak, \emph{Nature}, 1993,
  \textbf{365}, 141--143\relax
\mciteBstWouldAddEndPuncttrue
\mciteSetBstMidEndSepPunct{\mcitedefaultmidpunct}
{\mcitedefaultendpunct}{\mcitedefaultseppunct}\relax
\EndOfBibitem
\bibitem[Ishikawa \emph{et~al.}(2003)Ishikawa, Sugita, Ishikawa, Koshihara, and
  Kaizu]{ishikawa2003lanthanide}
N.~Ishikawa, M.~Sugita, T.~Ishikawa, S.-y. Koshihara and Y.~Kaizu,
  \emph{Journal of the American Chemical Society}, 2003, \textbf{125},
  8694--8695\relax
\mciteBstWouldAddEndPuncttrue
\mciteSetBstMidEndSepPunct{\mcitedefaultmidpunct}
{\mcitedefaultendpunct}{\mcitedefaultseppunct}\relax
\EndOfBibitem
\bibitem[Jiang \emph{et~al.}(2011)Jiang, Wang, Sun, Wang, and Gao]{jiang2011}
S.-D. Jiang, B.-W. Wang, H.-L. Sun, Z.-M. Wang and S.~Gao, \emph{Journal of the
  American Chemical Society}, 2011, \textbf{133}, 4730–--4733\relax
\mciteBstWouldAddEndPuncttrue
\mciteSetBstMidEndSepPunct{\mcitedefaultmidpunct}
{\mcitedefaultendpunct}{\mcitedefaultseppunct}\relax
\EndOfBibitem
\bibitem[Harriman \emph{et~al.}(2017)Harriman, Brosmer, Ungur, Diaconescu, and
  Murugesu]{Harriman2017}
K.~L. Harriman, J.~L. Brosmer, L.~Ungur, P.~L. Diaconescu and M.~Murugesu,
  \emph{Journal of the American Chemical Society}, 2017, \textbf{139},
  1420--1423\relax
\mciteBstWouldAddEndPuncttrue
\mciteSetBstMidEndSepPunct{\mcitedefaultmidpunct}
{\mcitedefaultendpunct}{\mcitedefaultseppunct}\relax
\EndOfBibitem
\bibitem[Guo \emph{et~al.}(2018)Guo, Day, Chen, Tong, Mansikkam{\"a}ki, and
  Layfield]{Guo2018}
F.-S. Guo, B.~M. Day, Y.-C. Chen, M.-L. Tong, A.~Mansikkam{\"a}ki and R.~A.
  Layfield, \emph{Science}, 2018, \textbf{362}, 1400--1403\relax
\mciteBstWouldAddEndPuncttrue
\mciteSetBstMidEndSepPunct{\mcitedefaultmidpunct}
{\mcitedefaultendpunct}{\mcitedefaultseppunct}\relax
\EndOfBibitem
\bibitem[Li \emph{et~al.}(2019)Li, Zhai, Chen, Ding, and Zheng]{Li2019}
Z.-H. Li, Y.-Q. Zhai, W.-P. Chen, Y.-S. Ding and Y.-Z. Zheng, \emph{Chemistry -
  A European Journal}, 2019, \textbf{25}, 16219--16224\relax
\mciteBstWouldAddEndPuncttrue
\mciteSetBstMidEndSepPunct{\mcitedefaultmidpunct}
{\mcitedefaultendpunct}{\mcitedefaultseppunct}\relax
\EndOfBibitem
\bibitem[Rinehart and Long(2011)]{Rinehart2011}
J.~D. Rinehart and J.~R. Long, \emph{Chemical Science}, 2011, \textbf{2},
  2078--2085\relax
\mciteBstWouldAddEndPuncttrue
\mciteSetBstMidEndSepPunct{\mcitedefaultmidpunct}
{\mcitedefaultendpunct}{\mcitedefaultseppunct}\relax
\EndOfBibitem
\bibitem[Ungur and Chibotaru(2016)]{Ungur2016}
L.~Ungur and L.~F. Chibotaru, \emph{Inorganic Chemistry}, 2016, \textbf{55},
  10043--10056\relax
\mciteBstWouldAddEndPuncttrue
\mciteSetBstMidEndSepPunct{\mcitedefaultmidpunct}
{\mcitedefaultendpunct}{\mcitedefaultseppunct}\relax
\EndOfBibitem
\bibitem[McAdams \emph{et~al.}(2017)McAdams, Ariciu, Kostopoulos, Walsh, and
  Tuna]{Mcadams2017}
S.~G. McAdams, A.-M. Ariciu, A.~K. Kostopoulos, J.~P. Walsh and F.~Tuna,
  \emph{Coordination Chemistry Reviews}, 2017, \textbf{346}, 216--239\relax
\mciteBstWouldAddEndPuncttrue
\mciteSetBstMidEndSepPunct{\mcitedefaultmidpunct}
{\mcitedefaultendpunct}{\mcitedefaultseppunct}\relax
\EndOfBibitem
\bibitem[Lunghi \emph{et~al.}(2017)Lunghi, Totti, Sessoli, and
  Sanvito]{Lunghi2017}
A.~Lunghi, F.~Totti, R.~Sessoli and S.~Sanvito, \emph{Nature Communications},
  2017, \textbf{8}, 14620\relax
\mciteBstWouldAddEndPuncttrue
\mciteSetBstMidEndSepPunct{\mcitedefaultmidpunct}
{\mcitedefaultendpunct}{\mcitedefaultseppunct}\relax
\EndOfBibitem
\bibitem[Ullah \emph{et~al.}(2019)Ullah, Cerd\'a, Baldov{\'\i}, Varganov,
  Arag\'o, and Gaita-Ari{\~n}o]{ullah2019silico}
A.~Ullah, J.~Cerd\'a, J.~J. Baldov{\'\i}, S.~A. Varganov, J.~Arag\'o and
  A.~Gaita-Ari{\~n}o, \emph{The Journal of Physical Chemistry Letters}, 2019,
  \textbf{10}, 7678--7683\relax
\mciteBstWouldAddEndPuncttrue
\mciteSetBstMidEndSepPunct{\mcitedefaultmidpunct}
{\mcitedefaultendpunct}{\mcitedefaultseppunct}\relax
\EndOfBibitem
\bibitem[Goodwin \emph{et~al.}(2017)Goodwin, Ortu, Reta, Chilton, and
  Mills]{Goodwin2017hysteresis}
C.~A. Goodwin, F.~Ortu, D.~Reta, N.~F. Chilton and D.~P. Mills, \emph{Nature},
  2017, \textbf{548}, 439--–442\relax
\mciteBstWouldAddEndPuncttrue
\mciteSetBstMidEndSepPunct{\mcitedefaultmidpunct}
{\mcitedefaultendpunct}{\mcitedefaultseppunct}\relax
\EndOfBibitem
\bibitem[McClain \emph{et~al.}(2018)McClain, Gould, Chakarawet, Teat, Groshens,
  Long, and Harvey]{McClain2018}
K.~R. McClain, C.~A. Gould, K.~Chakarawet, S.~J. Teat, T.~J. Groshens, J.~R.
  Long and B.~G. Harvey, \emph{Chemical Science}, 2018, \textbf{9},
  8492--8503\relax
\mciteBstWouldAddEndPuncttrue
\mciteSetBstMidEndSepPunct{\mcitedefaultmidpunct}
{\mcitedefaultendpunct}{\mcitedefaultseppunct}\relax
\EndOfBibitem
\bibitem[Atzori \emph{et~al.}(2016)Atzori, Morra, Tesi, Albino, Chiesa, Sorace,
  and Sessoli]{Atzori2016}
M.~Atzori, E.~Morra, L.~Tesi, A.~Albino, M.~Chiesa, L.~Sorace and R.~Sessoli,
  \emph{Journal of the American Chemical Society}, 2016, \textbf{138},
  11234--11244\relax
\mciteBstWouldAddEndPuncttrue
\mciteSetBstMidEndSepPunct{\mcitedefaultmidpunct}
{\mcitedefaultendpunct}{\mcitedefaultseppunct}\relax
\EndOfBibitem
\bibitem[Rosaleny \emph{et~al.}(2017)Rosaleny, Zinovjev, Tu\~n\'on, and
  Gaita-Ari\~no]{Rosaleny2019}
L.~E. Rosaleny, K.~Zinovjev, I.~Tu\~n\'on and A.~Gaita-Ari\~no, \emph{Physical
  Chemistry Chemical Physics}, 2017, \textbf{548}, 439--–442\relax
\mciteBstWouldAddEndPuncttrue
\mciteSetBstMidEndSepPunct{\mcitedefaultmidpunct}
{\mcitedefaultendpunct}{\mcitedefaultseppunct}\relax
\EndOfBibitem
\bibitem[O’Neal \emph{et~al.}(2019)O’Neal, Paul, Al-Wahish, Hughey,
  Blockmon, Luo, Cheong, Zapf, Topping, Singleton,\emph{et~al.}]{ONeal2019}
K.~R. O’Neal, A.~Paul, A.~Al-Wahish, K.~D. Hughey, A.~L. Blockmon, X.~Luo,
  S.-W. Cheong, V.~S. Zapf, C.~V. Topping, J.~Singleton \emph{et~al.},
  \emph{npj Quantum Materials}, 2019, \textbf{4}, 1--6\relax
\mciteBstWouldAddEndPuncttrue
\mciteSetBstMidEndSepPunct{\mcitedefaultmidpunct}
{\mcitedefaultendpunct}{\mcitedefaultseppunct}\relax
\EndOfBibitem
\bibitem[Giansiracusa \emph{et~al.}(2019)Giansiracusa, Kostopoulos, Collison,
  Winpenny, and Chilton]{Giansiracusa2019}
M.~J. Giansiracusa, A.~K. Kostopoulos, D.~Collison, R.~E.~P. Winpenny and N.~F.
  Chilton, \emph{Chemical Communications}, 2019, \textbf{55}, 7025--7028\relax
\mciteBstWouldAddEndPuncttrue
\mciteSetBstMidEndSepPunct{\mcitedefaultmidpunct}
{\mcitedefaultendpunct}{\mcitedefaultseppunct}\relax
\EndOfBibitem
\bibitem[Castro-Alvarez \emph{et~al.}(2020)Castro-Alvarez, Gil~Sanchez, Llanos,
  and Aravena]{Castro2020}
A.~Castro-Alvarez, Y.~Gil~Sanchez, L.~C. Llanos and D.~Aravena, \emph{Inorganic
  Chemistry Frontiers}, 2020, \textbf{7}, 2478--2486\relax
\mciteBstWouldAddEndPuncttrue
\mciteSetBstMidEndSepPunct{\mcitedefaultmidpunct}
{\mcitedefaultendpunct}{\mcitedefaultseppunct}\relax
\EndOfBibitem
\bibitem[Long \emph{et~al.}(2020)Long, Ivanov, Khomchenko, Mamontova, Thibaud,
  Rouquette, Beaudhuin, Granier, Ferreira, Carlos,\emph{et~al.}]{Long2020}
J.~Long, M.~S. Ivanov, V.~A. Khomchenko, E.~Mamontova, J.-M. Thibaud,
  J.~Rouquette, M.~Beaudhuin, D.~Granier, R.~A. Ferreira, L.~D. Carlos
  \emph{et~al.}, \emph{Science}, 2020, \textbf{367}, 671--676\relax
\mciteBstWouldAddEndPuncttrue
\mciteSetBstMidEndSepPunct{\mcitedefaultmidpunct}
{\mcitedefaultendpunct}{\mcitedefaultseppunct}\relax
\EndOfBibitem
\bibitem[Garlatti \emph{et~al.}(2020)Garlatti, Tesi, Lunghi, Atzori, Voneshen,
  Santini, Sanvito, Guidi, Sessoli, and Carretta]{garlatti2020unveiling}
E.~Garlatti, L.~Tesi, A.~Lunghi, M.~Atzori, D.~Voneshen, P.~Santini,
  S.~Sanvito, T.~Guidi, R.~Sessoli and S.~Carretta, \emph{Nature
  Communications}, 2020, \textbf{11}, 1--10\relax
\mciteBstWouldAddEndPuncttrue
\mciteSetBstMidEndSepPunct{\mcitedefaultmidpunct}
{\mcitedefaultendpunct}{\mcitedefaultseppunct}\relax
\EndOfBibitem
\bibitem[Lee \emph{et~al.}(2006)Lee, Hase, Sugawara, Yoshizawa, and
  Sato]{Lee2006}
C.~H. Lee, I.~Hase, H.~Sugawara, H.~Yoshizawa and H.~J. Sato, \emph{Phys. Soc.
  Jpn.}, 2006, \textbf{75}, 123602\relax
\mciteBstWouldAddEndPuncttrue
\mciteSetBstMidEndSepPunct{\mcitedefaultmidpunct}
{\mcitedefaultendpunct}{\mcitedefaultseppunct}\relax
\EndOfBibitem
\bibitem[Christensen \emph{et~al.}(2008)Christensen, Abrahamsen, Christensen,
  Juranyi, Andersen, Lefmann, Andreasson, Bahl, and Iversen]{Christensen2008}
M.~Christensen, A.~B. Abrahamsen, N.~B. Christensen, F.~Juranyi, N.~H.
  Andersen, K.~Lefmann, J.~Andreasson, C.~R.~H. Bahl and B.~B. Iversen,
  \emph{Nature Materials}, 2008, \textbf{7}, 811–--815\relax
\mciteBstWouldAddEndPuncttrue
\mciteSetBstMidEndSepPunct{\mcitedefaultmidpunct}
{\mcitedefaultendpunct}{\mcitedefaultseppunct}\relax
\EndOfBibitem
\bibitem[Toberer \emph{et~al.}(2011)Toberer, Zevalkink, and
  Snyder]{Toberer2011}
E.~S. Toberer, A.~Zevalkink and G.~J. Snyder, \emph{Journal of Materials
  Chemistry}, 2011, \textbf{21}, 15843--15852\relax
\mciteBstWouldAddEndPuncttrue
\mciteSetBstMidEndSepPunct{\mcitedefaultmidpunct}
{\mcitedefaultendpunct}{\mcitedefaultseppunct}\relax
\EndOfBibitem
\bibitem[Lunghi \emph{et~al.}(2017)Lunghi, Totti, Sanvito, and
  Sessoli]{Lunghi2017CS}
A.~Lunghi, F.~Totti, S.~Sanvito and R.~Sessoli, \emph{Chemical Science}, 2017,
  \textbf{8}, 6051–--6059\relax
\mciteBstWouldAddEndPuncttrue
\mciteSetBstMidEndSepPunct{\mcitedefaultmidpunct}
{\mcitedefaultendpunct}{\mcitedefaultseppunct}\relax
\EndOfBibitem
\bibitem[Shiddiq \emph{et~al.}(2016)Shiddiq, Komijani, Duan, Gaita-Ari{\~n}o,
  Coronado, and Hill]{Shiddiq2016}
M.~Shiddiq, D.~Komijani, Y.~Duan, A.~Gaita-Ari{\~n}o, E.~Coronado and S.~Hill,
  \emph{Nature}, 2016, \textbf{531}, 348--351\relax
\mciteBstWouldAddEndPuncttrue
\mciteSetBstMidEndSepPunct{\mcitedefaultmidpunct}
{\mcitedefaultendpunct}{\mcitedefaultseppunct}\relax
\EndOfBibitem
\bibitem[Shiozaki \emph{et~al.}(1996)Shiozaki, Inagaki, Nishino, Nishio,
  Maekawa, Kominami, and Kera]{Shiozaki1996}
R.~Shiozaki, A.~Inagaki, A.~Nishino, E.~Nishio, M.~Maekawa, H.~Kominami and
  Y.~Kera, \emph{Journal of Alloys and Compounds}, 1996, \textbf{234},
  193--198\relax
\mciteBstWouldAddEndPuncttrue
\mciteSetBstMidEndSepPunct{\mcitedefaultmidpunct}
{\mcitedefaultendpunct}{\mcitedefaultseppunct}\relax
\EndOfBibitem
\bibitem[AlDamen \emph{et~al.}(2009)AlDamen, Cardona-Serra, Clemente-Juan,
  Coronado, Gaita-Ari{\~n}o, Mart{\'\i}-Gastaldo, Luis, and
  Montero]{AlDamen2009}
M.~A. AlDamen, S.~Cardona-Serra, J.~M. Clemente-Juan, E.~Coronado,
  A.~Gaita-Ari{\~n}o, C.~Mart{\'\i}-Gastaldo, F.~Luis and O.~Montero,
  \emph{Inorganic Chemistry}, 2009, \textbf{48}, 3467--3479\relax
\mciteBstWouldAddEndPuncttrue
\mciteSetBstMidEndSepPunct{\mcitedefaultmidpunct}
{\mcitedefaultendpunct}{\mcitedefaultseppunct}\relax
\EndOfBibitem
\bibitem[Ghosh \emph{et~al.}(2012)Ghosh, Datta, Friend, Cardona-Serra,
  Gaita-Ari{\~n}o, Coronado, and Hill]{Ghosh2012}
S.~Ghosh, S.~Datta, L.~Friend, S.~Cardona-Serra, A.~Gaita-Ari{\~n}o,
  E.~Coronado and S.~Hill, \emph{Dalton Transactions}, 2012, \textbf{41},
  13697--13704\relax
\mciteBstWouldAddEndPuncttrue
\mciteSetBstMidEndSepPunct{\mcitedefaultmidpunct}
{\mcitedefaultendpunct}{\mcitedefaultseppunct}\relax
\EndOfBibitem
\bibitem[Vonci \emph{et~al.}(2017)Vonci, Giansiracusa, Van~den Heuvel, Gable,
  Moubaraki, Murray, Yu, Mole, Soncini, and Boskovic]{Vonci2017}
M.~Vonci, M.~J. Giansiracusa, W.~Van~den Heuvel, R.~W. Gable, B.~Moubaraki,
  K.~S. Murray, D.~Yu, R.~A. Mole, A.~Soncini and C.~Boskovic, \emph{Inorganic
  Chemistry}, 2017, \textbf{56}, 378--394\relax
\mciteBstWouldAddEndPuncttrue
\mciteSetBstMidEndSepPunct{\mcitedefaultmidpunct}
{\mcitedefaultendpunct}{\mcitedefaultseppunct}\relax
\EndOfBibitem
\bibitem[Escalera-Moreno and Baldov{\'\i}(2019)]{escalera2019unveiling}
L.~Escalera-Moreno and J.~J. Baldov{\'\i}, \emph{Frontiers in chemistry}, 2019,
  \textbf{7}, 662\relax
\mciteBstWouldAddEndPuncttrue
\mciteSetBstMidEndSepPunct{\mcitedefaultmidpunct}
{\mcitedefaultendpunct}{\mcitedefaultseppunct}\relax
\EndOfBibitem
\bibitem[Liu \emph{et~al.}(2021)Liu, Mrozek, Duan, Ullah, Baldoví, Coronado,
  Gaita-Ariño, and Ardavan]{Liu2021}
J.~Liu, J.~Mrozek, Y.~Duan, A.~Ullah, J.~J. Baldoví, E.~Coronado,
  A.~Gaita-Ariño and A.~Ardavan, \emph{arXiv}, 2021,  2005.01029\relax
\mciteBstWouldAddEndPuncttrue
\mciteSetBstMidEndSepPunct{\mcitedefaultmidpunct}
{\mcitedefaultendpunct}{\mcitedefaultseppunct}\relax
\EndOfBibitem
\bibitem[Blockmon \emph{et~al.}(2021)Blockmon, Ullah, Hughey, Duan, O'Neal,
  Ozerov, Baldoví, Arag\'o, Gaita-Ariño, Coronado, and
  Musfeldt]{Blockmon2021}
A.~L. Blockmon, A.~Ullah, K.~D. Hughey, Y.~Duan, K.~R. O'Neal, M.~Ozerov, J.~J.
  Baldoví, J.~Arag\'o, A.~Gaita-Ariño, E.~Coronado and J.~L. Musfeldt,
  \emph{arXiv}, 2021,  2102.08713\relax
\mciteBstWouldAddEndPuncttrue
\mciteSetBstMidEndSepPunct{\mcitedefaultmidpunct}
{\mcitedefaultendpunct}{\mcitedefaultseppunct}\relax
\EndOfBibitem
\bibitem[Gaita-Ari\~no and Schechter(2011)]{Gaita2011}
A.~Gaita-Ari\~no and M.~Schechter, \emph{Physical Review Letters}, 2011,
  \textbf{107}, 105504\relax
\mciteBstWouldAddEndPuncttrue
\mciteSetBstMidEndSepPunct{\mcitedefaultmidpunct}
{\mcitedefaultendpunct}{\mcitedefaultseppunct}\relax
\EndOfBibitem
\bibitem[Natterer \emph{et~al.}(2017)Natterer, Yang, Paul, Willke, Choi,
  Greber, Heinrich, and Lutz]{Natterer2017}
F.~B. Natterer, K.~Yang, W.~Paul, P.~Willke, T.~Choi, T.~Greber, A.~J. Heinrich
  and C.~P. Lutz, \emph{Nature}, 2017, \textbf{543}, 226--228\relax
\mciteBstWouldAddEndPuncttrue
\mciteSetBstMidEndSepPunct{\mcitedefaultmidpunct}
{\mcitedefaultendpunct}{\mcitedefaultseppunct}\relax
\EndOfBibitem
\bibitem[Donati \emph{et~al.}(2020)Donati, Rusponi, Stepanow, Persichetti,
  Singha, Juraschek, Wäckerlin, Baltic, Pivetta, Diller, Nistor, Dreiser,
  Kummer, Velez-Fort, Spaldin, Brune, and Gambardella]{Donati2020}
F.~Donati, S.~Rusponi, S.~Stepanow, L.~Persichetti, A.~Singha, D.~M. Juraschek,
  C.~Wäckerlin, R.~Baltic, M.~Pivetta, K.~Diller, C.~Nistor, J.~Dreiser,
  K.~Kummer, E.~Velez-Fort, N.~A. Spaldin, H.~Brune and P.~Gambardella,
  \emph{Physical Review Letters}, 2020, \textbf{124}, 077224\relax
\mciteBstWouldAddEndPuncttrue
\mciteSetBstMidEndSepPunct{\mcitedefaultmidpunct}
{\mcitedefaultendpunct}{\mcitedefaultseppunct}\relax
\EndOfBibitem
\bibitem[Shpakov \emph{et~al.}(2005)Shpakov, Gotte, Baudin, Woo, and
  Hermansson]{Shpakov2005}
V.~Shpakov, A.~Gotte, M.~Baudin, T.~Woo and K.~Hermansson, \emph{Physical
  Reviews B}, 2005, \textbf{72}, 195427\relax
\mciteBstWouldAddEndPuncttrue
\mciteSetBstMidEndSepPunct{\mcitedefaultmidpunct}
{\mcitedefaultendpunct}{\mcitedefaultseppunct}\relax
\EndOfBibitem
\bibitem[Baldov{\'\i} \emph{et~al.}(2016)Baldov{\'\i}, Duan, Bustos,
  Cardona-Serra, Gouzerh, Villanneau, Gontard, Clemente-Juan, Gaita-Ari{\~n}o,
  Gim{\'e}nez-Saiz,\emph{et~al.}]{baldovi2016single}
J.~J. Baldov{\'\i}, Y.~Duan, C.~Bustos, S.~Cardona-Serra, P.~Gouzerh,
  R.~Villanneau, G.~Gontard, J.~M. Clemente-Juan, A.~Gaita-Ari{\~n}o,
  C.~Gim{\'e}nez-Saiz \emph{et~al.}, \emph{Dalton Transactions}, 2016,
  \textbf{45}, 16653--16660\relax
\mciteBstWouldAddEndPuncttrue
\mciteSetBstMidEndSepPunct{\mcitedefaultmidpunct}
{\mcitedefaultendpunct}{\mcitedefaultseppunct}\relax
\EndOfBibitem
\bibitem[Stuckart \emph{et~al.}(2018)Stuckart, Izarova, van Leusen, Smekhova,
  Schmitz-Antoniak, Bamberger, van Slageren, Santiago-Sch{\"u}bel, and
  K{\"o}gerler]{stuckart2018host}
M.~Stuckart, N.~V. Izarova, J.~van Leusen, A.~Smekhova, C.~Schmitz-Antoniak,
  H.~Bamberger, J.~van Slageren, B.~Santiago-Sch{\"u}bel and P.~K{\"o}gerler,
  \emph{Chemistry--A European Journal}, 2018, \textbf{24}, 17767--17778\relax
\mciteBstWouldAddEndPuncttrue
\mciteSetBstMidEndSepPunct{\mcitedefaultmidpunct}
{\mcitedefaultendpunct}{\mcitedefaultseppunct}\relax
\EndOfBibitem
\bibitem[Vance \emph{et~al.}(2003)Vance, Willey, van Buuren, Nelson, Bostedt,
  Fox, and Terminello]{vance2003xas}
A.~L. Vance, T.~M. Willey, T.~van Buuren, A.~Nelson, C.~Bostedt, G.~A. Fox and
  L.~J. Terminello, \emph{Nano Letters}, 2003, \textbf{3}, 81--84\relax
\mciteBstWouldAddEndPuncttrue
\mciteSetBstMidEndSepPunct{\mcitedefaultmidpunct}
{\mcitedefaultendpunct}{\mcitedefaultseppunct}\relax
\EndOfBibitem
\bibitem[Konstantinov \emph{et~al.}(2021)Konstantinov, Tauzin, Noumb{\'e},
  Dragoe, Kundys, Majjad, Brosseau, Lenertz, Singh,
  Berciaud,\emph{et~al.}]{konstantinov2021}
N.~Konstantinov, A.~Tauzin, U.~N. Noumb{\'e}, D.~Dragoe, B.~Kundys, H.~Majjad,
  A.~Brosseau, M.~Lenertz, A.~Singh, S.~Berciaud \emph{et~al.}, \emph{Journal
  of Materials Chemistry C}, 2021, \textbf{9}, 2712--2720\relax
\mciteBstWouldAddEndPuncttrue
\mciteSetBstMidEndSepPunct{\mcitedefaultmidpunct}
{\mcitedefaultendpunct}{\mcitedefaultseppunct}\relax
\EndOfBibitem
\bibitem[Boix-Constant \emph{et~al.}(2020)Boix-Constant, Mañas-Valero,
  Córdoba, Baldoví, Rubio, and Coronado]{boix2020}
C.~Boix-Constant, S.~Mañas-Valero, R.~Córdoba, J.~J. Baldoví, A.~Rubio and
  E.~Coronado, \emph{arXiv preprint arXiv:2009.14550}, 2020\relax
\mciteBstWouldAddEndPuncttrue
\mciteSetBstMidEndSepPunct{\mcitedefaultmidpunct}
{\mcitedefaultendpunct}{\mcitedefaultseppunct}\relax
\EndOfBibitem
\end{mcitethebibliography}
\bibliographystyle{rsc}

\newpage
\clearpage

\end{document}


\onecolumn
\section*{Supplementary Tables}

\begin{longtable}{c|c|c|c|c|c}
\caption{Frequencies (cm$^{-1}$), vibronic coupling (S$_n$ (cm$^{-1}$) at CASSCF level for optimized geometry for each normal mode ($n$) and vibronic-phonon couplings $S(\mu_{n,xx})$, $S(\mu_{n,xy})$, $S(\mu_{n,xz})$. }
\label{tab:IR-OPT-cou}\\
\hline
    \multicolumn{1}{p{5em}|}{\centering Modes   (n)} & \multicolumn{1}{p{5.0em}|}{\centering Frequencies (cm$^{-1}$)} &    \multicolumn{1}{p{5em}|}{\centering S$_n$   (cm$^{-1}$)} & \multicolumn{1}{p{5em}|}{\centering S  (n, $\mu_{xx}$)}  &
    \multicolumn{1}{p{5em}|}{\centering S  (n, $\mu_{xy}$)} &
    \multicolumn{1}{p{5em}}{\centering S  (n, $\mu_{xz}$)}\\
\hline
\endfirsthead

\multicolumn{3}{c}

{{\tablename\ \thetable{}: Continued from previous page}}\\
\hline
    \multicolumn{1}{p{5em}|}{\centering Modes   (n)} & \multicolumn{1}{p{5.0em}|}{\centering Frequencies (cm$^{-1}$)} &    \multicolumn{1}{p{5em}|}{\centering S$_n$   (cm$^{-1}$)} & \multicolumn{1}{p{5em}|}{\centering S  (n, $\mu_{xx}$)}  &
    \multicolumn{1}{p{5em}|}{\centering S  (n, $\mu_{xy}$)} &
    \multicolumn{1}{p{5em}}{\centering S  (n, $\mu_{xz}$)}\\
    \hline
    \endhead
    \hline
\endlastfoot

    1     & 8.62  & 0.085 & 0.01  & 0.01  & 0.01 \\
    2     & 17.94 & 0.176 & 0.14  & 0.03  & 0.01 \\
    3     & 18.43 & 0.119 & 0.03  & 0.04  & 0.01 \\
    4     & 24.33 & 0.247 & 0.00  & 0.13  & 0.01 \\
    5     & 29.97 & 0.220 & 0.01  & 0.01  & 0.13 \\
    6     & 43.22 & 0.123 & 0.91  & 0.01  & 0.01 \\
    7     & 54.02 & 0.150 & 0.26  & 0.03  & 0.00 \\
    8     & 58.39 & 0.180 & 0.01  & 0.01  & 0.03 \\
    9     & 67.12 & 0.145 & 0.17  & 0.01  & 0.00 \\
    10    & 70.95 & 0.130 & 0.04  & 0.05  & 0.01 \\
    11    & 99.61 & 0.125 & 0.01  & 0.07  & 0.01 \\
    12    & 106.52 & 0.151 & 0.02  & 0.01  & 0.07 \\
    13    & 106.78 & 0.051 & 0.01  & 0.01  & 0.04 \\
    14    & 107.26 & 0.132 & 0.00  & 0.06  & 0.01 \\
    15    & 108.76 & 0.167 & 0.01  & 0.00  & 0.00 \\
    16    & 114.68 & 0.067 & 0.01  & 0.00  & 0.00 \\
    17    & 114.71 & 0.080 & 0.01  & 0.00  & 0.00 \\
    18    & 140.11 & 0.109 & 0.01  & 0.01  & 0.00 \\
    19    & 140.76 & 0.040 & 0.04  & 0.00  & 0.01 \\
    20    & 141.19 & 0.038 & 0.00  & 0.01  & 0.04 \\
    21    & 142.11 & 0.129 & 0.01  & 0.04  & 0.00 \\
    22    & 143.98 & 0.123 & 0.09  & 0.01  & 0.00 \\
    23    & 144.75 & 0.039 & 0.02  & 0.06  & 0.00 \\
    24    & 157.60 & 0.020 & 0.00  & 0.00  & 0.02 \\
    25    & 158.17 & 0.081 & 0.00  & 0.02  & 0.23 \\
    26    & 159.14 & 0.096 & 0.00  & 0.28  & 0.02 \\
    27    & 159.88 & 0.042 & 0.00  & 0.08  & 0.03 \\
    28    & 160.56 & 0.039 & 0.00  & 0.03  & 0.12 \\
    29    & 161.30 & 0.079 & 0.01  & 0.06  & 0.08 \\
    30    & 167.76 & 0.094 & 0.00  & 0.00  & 0.00 \\
    31    & 167.77 & 0.083 & 0.01  & 0.02  & 0.00 \\
    32    & 184.62 & 0.122 & 0.00  & 0.00  & 0.01 \\
    33    & 184.73 & 0.122 & 0.00  & 0.01  & 0.00 \\
    34    & 193.02 & 0.038 & 0.01  & 0.01  & 0.01 \\
    35    & 202.14 & 0.095 & 0.15  & 0.00  & 0.00 \\
    36    & 216.40 & 0.027 & 0.02  & 0.02  & 0.01 \\
    37    & 216.89 & 0.066 & 0.00  & 0.02  & 0.03 \\
    38    & 218.12 & 0.088 & 0.00  & 0.00  & 0.06 \\
    39    & 218.13 & 0.099 & 0.01  & 0.02  & 0.06 \\
    40    & 218.22 & 0.126 & 0.01  & 0.11  & 0.08 \\
    41    & 218.24 & 0.117 & 0.00  & 0.11  & 0.07 \\
    42    & 219.96 & 0.057 & 0.01  & 0.01  & 0.02 \\
    43    & 220.00 & 0.054 & 0.00  & 0.14  & 0.02 \\
    44    & 220.07 & 0.115 & 0.00  & 0.01  & 0.18 \\
    45    & 220.47 & 0.111 & 0.03  & 0.05  & 0.02 \\
    46    & 223.18 & 0.079 & 0.05  & 0.08  & 0.01 \\
    47    & 224.41 & 0.072 & 0.05  & 0.04  & 0.01 \\
    48    & 230.85 & 0.034 & 0.01  & 0.01  & 0.00 \\
    49    & 231.24 & 0.031 & 0.01  & 0.01  & 0.02 \\
    50    & 231.75 & 0.036 & 0.00  & 0.02  & 0.02 \\
    51    & 232.10 & 0.036 & 0.01  & 0.02  & 0.00 \\
    52    & 235.13 & 0.019 & 0.00  & 0.00  & 0.01 \\
    53    & 235.33 & 0.029 & 0.01  & 0.00  & 0.00 \\
    54    & 257.49 & 0.097 & 0.00  & 0.00  & 0.00 \\
    55    & 257.49 & 0.108 & 0.01  & 0.00  & 0.00 \\
    56    & 289.55 & 0.030 & 0.00  & 0.00  & 0.00 \\
    57    & 291.07 & 0.025 & 0.00  & 0.00  & 0.00 \\
    58    & 329.19 & 0.091 & 0.00  & 0.00  & 0.02 \\
    59    & 329.25 & 0.125 & 0.00  & 0.02  & 0.01 \\
    60    & 337.26 & 0.101 & 0.00  & 0.02  & 0.02 \\
    61    & 337.36 & 0.143 & 0.00  & 0.02  & 0.04 \\
    62    & 337.40 & 0.099 & 0.00  & 0.03  & 0.00 \\
    63    & 337.72 & 0.101 & 0.00  & 0.02  & 0.01 \\
    64    & 351.19 & 0.101 & 0.00  & 0.08  & 0.03 \\
    65    & 354.95 & 0.147 & 0.00  & 0.07  & 0.21 \\
    66    & 355.21 & 0.148 & 0.00  & 0.20  & 0.08 \\
    67    & 359.02 & 0.064 & 0.01  & 0.00  & 0.00 \\
    68    & 360.26 & 0.032 & 0.02  & 0.01  & 0.00 \\
    69    & 363.49 & 0.085 & 0.00  & 0.04  & 0.02 \\
    70    & 399.03 & 0.041 & 0.02  & 0.05  & 0.06 \\
    71    & 399.78 & 0.067 & 0.01  & 0.07  & 0.05 \\
    72    & 399.80 & 0.053 & 0.02  & 0.01  & 0.03 \\
    73    & 404.19 & 0.083 & 0.30  & 0.00  & 0.00 \\
    74    & 414.55 & 0.141 & 0.01  & 0.26  & 0.03 \\
    75    & 414.92 & 0.126 & 0.00  & 0.04  & 0.27 \\
    76    & 416.67 & 0.116 & 0.00  & 0.07  & 0.18 \\
    77    & 416.90 & 0.111 & 0.04  & 0.14  & 0.06 \\
    78    & 417.53 & 0.131 & 0.02  & 0.10  & 0.06 \\
    79    & 417.69 & 0.120 & 0.02  & 0.06  & 0.04 \\
    80    & 429.95 & 0.091 & 0.00  & 0.03  & 0.00 \\
    81    & 431.15 & 0.065 & 0.35  & 0.01  & 0.00 \\
    82    & 435.19 & 0.083 & 0.01  & 0.01  & 0.17 \\
    83    & 436.31 & 0.046 & 0.03  & 0.05  & 0.05 \\
    84    & 436.72 & 0.078 & 0.07  & 0.15  & 0.03 \\
    85    & 437.91 & 0.056 & 0.15  & 0.08  & 0.00 \\
    86    & 479.05 & 0.084 & 0.00  & 0.21  & 0.01 \\
    87    & 479.34 & 0.070 & 0.00  & 0.01  & 0.21 \\
    88    & 480.96 & 0.037 & 0.01  & 0.01  & 0.00 \\
    89    & 481.29 & 0.038 & 0.00  & 0.00  & 0.03 \\
    90    & 489.34 & 0.055 & 0.00  & 0.02  & 0.01 \\
    91    & 489.38 & 0.054 & 0.00  & 0.00  & 0.02 \\
    92    & 514.29 & 0.028 & 0.00  & 0.00  & 0.00 \\
    93    & 515.09 & 0.031 & 0.01  & 0.00  & 0.01 \\
    94    & 515.25 & 0.029 & 0.02  & 0.00  & 0.01 \\
    95    & 516.21 & 0.076 & 0.12  & 0.00  & 0.00 \\
    96    & 520.31 & 0.113 & 0.00  & 0.01  & 0.01 \\
    97    & 521.09 & 0.102 & 0.01  & 0.01  & 0.01 \\
    98    & 528.07 & 0.086 & 0.04  & 0.01  & 0.00 \\
    99    & 529.22 & 0.079 & 0.01  & 0.00  & 0.01 \\
    100   & 530.46 & 0.069 & 0.09  & 0.00  & 0.00 \\
    101   & 531.81 & 0.080 & 0.19  & 0.01  & 0.00 \\
    102   & 535.59 & 0.091 & 0.00  & 0.00  & 0.01 \\
    103   & 536.21 & 0.097 & 0.05  & 0.02  & 0.00 \\
    104   & 554.98 & 0.020 & 0.00  & 0.00  & 0.00 \\
    105   & 555.04 & 0.018 & 0.00  & 0.00  & 0.00 \\
    106   & 578.43 & 0.031 & 0.01  & 0.14  & 0.01 \\
    107   & 579.04 & 0.042 & 0.01  & 0.06  & 0.11 \\
    108   & 579.13 & 0.027 & 0.02  & 0.14  & 0.01 \\
    109   & 579.49 & 0.033 & 0.00  & 0.02  & 0.18 \\
    110   & 671.58 & 0.235 & 0.00  & 0.25  & 0.01 \\
 \hline
\end{longtable}

\begin{longtable}{c|c|c|c|c|c|c}
\caption{For normal mode n, vibronic-phonon couplings $S(\mu_{n,yx})$, $S(\mu_{n,yy})$, $S(\mu_{n,yz})$, $S(\mu_{n,zx})$, $S(\mu_{n,zy})$, $S(\mu_{n,zz})$.}
\label{tab:IR-OPT-cou}\\
\hline
    \multicolumn{1}{p{5em}|}{\centering Modes   (n)} & \multicolumn{1}{p{5em}|}{\centering S  (n, $\mu_{yx}$)} &    \multicolumn{1}{p{5em}|}{\centering S  (n, $\mu_{yy}$)} & \multicolumn{1}{p{5em}|}{\centering S  (n, $\mu_{yz}$)}  &
    \multicolumn{1}{p{5em}|}{\centering S  (n, $\mu_{zx}$)} &
    \multicolumn{1}{p{5em}|}{\centering S  (n, $\mu_{zy}$)} &
    \multicolumn{1}{p{5em}}{\centering S  (n, $\mu_{zz}$)} \\
\hline
\endfirsthead

\multicolumn{3}{c}

{{\tablename\ \thetable{}: Continued from previous page}}\\
\hline
    \multicolumn{1}{p{5em}|}{\centering Modes   (n)} & \multicolumn{1}{p{5em}|}{\centering S  (n, $\mu_{yx}$)} &    \multicolumn{1}{p{5em}|}{\centering S  (n, $\mu_{yy}$)} & \multicolumn{1}{p{5em}|}{\centering S  (n, $\mu_{yz}$)}  &
    \multicolumn{1}{p{5em}|}{\centering S  (n, $\mu_{zx}$)} &
    \multicolumn{1}{p{5em}|}{\centering S  (n, $\mu_{zy}$)} &
    \multicolumn{1}{p{5em}}{\centering S  (n, $\mu_{zz}$)} \\
    \hline
    \endhead
    \hline
\endlastfoot
    1     & 0.03  & 0.01  & 0.00  & 0.04  & 0.00  & 0.00 \\
    2     & 0.13  & 0.03  & 0.00  & 0.02  & 0.01  & 0.04 \\
    3     & 0.21  & 0.01  & 0.02  & 0.04  & 0.02  & 0.01 \\
    4     & 0.65  & 0.05  & 0.01  & 0.05  & 0.01  & 0.05 \\
    5     & 0.06  & 0.01  & 0.04  & 0.69  & 0.05  & 0.01 \\
    6     & 0.06  & 0.00  & 0.01  & 0.00  & 0.00  & 0.06 \\
    7     & 0.14  & 0.31  & 0.02  & 0.01  & 0.02  & 0.25 \\
    8     & 0.02  & 0.03  & 0.33  & 0.11  & 0.34  & 0.02 \\
    9     & 0.05  & 0.20  & 0.01  & 0.01  & 0.01  & 0.28 \\
    10    & 0.03  & 0.03  & 0.07  & 0.01  & 0.08  & 0.03 \\
    11    & 0.32  & 0.00  & 0.01  & 0.04  & 0.01  & 0.00 \\
    12    & 0.04  & 0.01  & 0.01  & 0.41  & 0.01  & 0.01 \\
    13    & 0.06  & 0.01  & 0.01  & 0.27  & 0.01  & 0.01 \\
    14    & 0.43  & 0.00  & 0.00  & 0.05  & 0.02  & 0.00 \\
    15    & 0.08  & 0.07  & 0.06  & 0.01  & 0.08  & 0.07 \\
    16    & 0.03  & 0.13  & 0.15  & 0.00  & 0.14  & 0.13 \\
    17    & 0.01  & 0.10  & 0.10  & 0.03  & 0.08  & 0.08 \\
    18    & 0.12  & 0.02  & 0.02  & 0.00  & 0.02  & 0.02 \\
    19    & 0.01  & 0.12  & 0.01  & 0.00  & 0.00  & 0.14 \\
    20    & 0.01  & 0.00  & 0.01  & 0.18  & 0.00  & 0.00 \\
    21    & 0.17  & 0.05  & 0.04  & 0.02  & 0.01  & 0.06 \\
    22    & 0.03  & 0.38  & 0.00  & 0.00  & 0.01  & 0.37 \\
    23    & 0.02  & 0.05  & 0.08  & 0.04  & 0.01  & 0.03 \\
    24    & 0.01  & 0.00  & 0.45  & 0.02  & 0.45  & 0.01 \\
    25    & 0.02  & 0.00  & 0.03  & 0.20  & 0.04  & 0.01 \\
    26    & 0.23  & 0.01  & 0.06  & 0.01  & 0.04  & 0.01 \\
    27    & 0.06  & 0.05  & 0.14  & 0.02  & 0.11  & 0.03 \\
    28    & 0.02  & 0.01  & 0.17  & 0.09  & 0.18  & 0.00 \\
    29    & 0.03  & 0.03  & 0.15  & 0.07  & 0.21  & 0.01 \\
    30    & 0.01  & 0.35  & 0.19  & 0.01  & 0.19  & 0.35 \\
    31    & 0.02  & 0.19  & 0.38  & 0.01  & 0.33  & 0.19 \\
    32    & 0.00  & 0.00  & 0.01  & 0.01  & 0.01  & 0.00 \\
    33    & 0.01  & 0.03  & 0.01  & 0.00  & 0.00  & 0.01 \\
    34    & 0.02  & 0.01  & 0.00  & 0.00  & 0.00  & 0.01 \\
    35    & 0.00  & 0.17  & 0.02  & 0.00  & 0.01  & 0.19 \\
    36    & 0.01  & 0.04  & 0.02  & 0.00  & 0.01  & 0.06 \\
    37    & 0.03  & 0.01  & 0.09  & 0.01  & 0.08  & 0.04 \\
    38    & 0.00  & 0.15  & 0.17  & 0.03  & 0.16  & 0.16 \\
    39    & 0.00  & 0.18  & 0.12  & 0.03  & 0.12  & 0.17 \\
    40    & 0.04  & 0.04  & 0.05  & 0.03  & 0.05  & 0.05 \\
    41    & 0.06  & 0.06  & 0.09  & 0.02  & 0.09  & 0.03 \\
    42    & 0.02  & 0.05  & 0.01  & 0.02  & 0.01  & 0.01 \\
    43    & 0.16  & 0.02  & 0.01  & 0.02  & 0.01  & 0.03 \\
    44    & 0.01  & 0.01  & 0.02  & 0.20  & 0.02  & 0.00 \\
    45    & 0.06  & 0.10  & 0.03  & 0.01  & 0.05  & 0.06 \\
    46    & 0.09  & 0.10  & 0.01  & 0.01  & 0.02  & 0.09 \\
    47    & 0.07  & 0.07  & 0.02  & 0.01  & 0.04  & 0.10 \\
    48    & 0.00  & 0.01  & 0.01  & 0.01  & 0.00  & 0.00 \\
    49    & 0.00  & 0.00  & 0.00  & 0.02  & 0.00  & 0.01 \\
    50    & 0.00  & 0.01  & 0.00  & 0.02  & 0.02  & 0.01 \\
    51    & 0.04  & 0.00  & 0.01  & 0.00  & 0.01  & 0.01 \\
    52    & 0.00  & 0.00  & 0.18  & 0.00  & 0.18  & 0.00 \\
    53    & 0.01  & 0.00  & 0.11  & 0.00  & 0.11  & 0.00 \\
    54    & 0.00  & 0.01  & 0.00  & 0.00  & 0.01  & 0.01 \\
    55    & 0.02  & 0.00  & 0.01  & 0.01  & 0.01  & 0.00 \\
    56    & 0.00  & 0.00  & 0.02  & 0.00  & 0.01  & 0.00 \\
    57    & 0.01  & 0.00  & 0.00  & 0.00  & 0.00  & 0.00 \\
    58    & 0.00  & 0.01  & 0.10  & 0.02  & 0.10  & 0.01 \\
    59    & 0.03  & 0.10  & 0.01  & 0.01  & 0.02  & 0.10 \\
    60    & 0.03  & 0.01  & 0.01  & 0.03  & 0.00  & 0.00 \\
    61    & 0.02  & 0.00  & 0.00  & 0.05  & 0.00  & 0.00 \\
    62    & 0.05  & 0.00  & 0.00  & 0.01  & 0.00  & 0.00 \\
    63    & 0.01  & 0.00  & 0.01  & 0.04  & 0.01  & 0.01 \\
    64    & 0.14  & 0.01  & 0.00  & 0.06  & 0.00  & 0.00 \\
    65    & 0.12  & 0.00  & 0.01  & 0.39  & 0.01  & 0.00 \\
    66    & 0.37  & 0.00  & 0.01  & 0.15  & 0.00  & 0.00 \\
    67    & 0.01  & 0.05  & 0.00  & 0.00  & 0.01  & 0.05 \\
    68    & 0.02  & 0.05  & 0.00  & 0.01  & 0.00  & 0.06 \\
    69    & 0.08  & 0.02  & 0.01  & 0.04  & 0.02  & 0.02 \\
    70    & 0.02  & 0.02  & 0.04  & 0.02  & 0.03  & 0.03 \\
    71    & 0.03  & 0.02  & 0.02  & 0.02  & 0.02  & 0.03 \\
    72    & 0.00  & 0.01  & 0.01  & 0.00  & 0.00  & 0.03 \\
    73    & 0.00  & 0.12  & 0.00  & 0.00  & 0.00  & 0.11 \\
    74    & 0.02  & 0.10  & 0.00  & 0.00  & 0.00  & 0.09 \\
    75    & 0.00  & 0.01  & 0.09  & 0.01  & 0.09  & 0.01 \\
    76    & 0.00  & 0.02  & 0.06  & 0.00  & 0.06  & 0.01 \\
    77    & 0.01  & 0.00  & 0.03  & 0.01  & 0.03  & 0.01 \\
    78    & 0.01  & 0.16  & 0.05  & 0.01  & 0.05  & 0.17 \\
    79    & 0.02  & 0.05  & 0.16  & 0.01  & 0.16  & 0.05 \\
    80    & 0.12  & 0.01  & 0.00  & 0.00  & 0.01  & 0.01 \\
    81    & 0.01  & 0.05  & 0.01  & 0.00  & 0.01  & 0.06 \\
    82    & 0.01  & 0.00  & 0.01  & 0.31  & 0.02  & 0.00 \\
    83    & 0.09  & 0.01  & 0.01  & 0.07  & 0.01  & 0.00 \\
    84    & 0.25  & 0.02  & 0.01  & 0.04  & 0.01  & 0.00 \\
    85    & 0.12  & 0.01  & 0.00  & 0.01  & 0.00  & 0.03 \\
    86    & 0.06  & 0.01  & 0.00  & 0.00  & 0.00  & 0.01 \\
    87    & 0.00  & 0.00  & 0.01  & 0.05  & 0.01  & 0.00 \\
    88    & 0.01  & 0.01  & 0.00  & 0.00  & 0.00  & 0.02 \\
    89    & 0.00  & 0.00  & 0.01  & 0.00  & 0.01  & 0.00 \\
    90    & 0.01  & 0.06  & 0.00  & 0.00  & 0.00  & 0.06 \\
    91    & 0.00  & 0.00  & 0.06  & 0.01  & 0.06  & 0.00 \\
    92    & 0.00  & 0.00  & 0.01  & 0.00  & 0.01  & 0.01 \\
    93    & 0.00  & 0.02  & 0.04  & 0.00  & 0.05  & 0.02 \\
    94    & 0.00  & 0.04  & 0.02  & 0.01  & 0.03  & 0.04 \\
    95    & 0.00  & 0.24  & 0.00  & 0.00  & 0.01  & 0.23 \\
    96    & 0.02  & 0.04  & 0.04  & 0.01  & 0.03  & 0.04 \\
    97    & 0.01  & 0.03  & 0.04  & 0.01  & 0.04  & 0.05 \\
    98    & 0.01  & 0.04  & 0.00  & 0.00  & 0.00  & 0.04 \\
    99    & 0.00  & 0.01  & 0.00  & 0.01  & 0.00  & 0.01 \\
    100   & 0.01  & 0.06  & 0.00  & 0.00  & 0.00  & 0.08 \\
    101   & 0.00  & 0.13  & 0.00  & 0.00  & 0.00  & 0.14 \\
    102   & 0.00  & 0.00  & 0.05  & 0.00  & 0.05  & 0.01 \\
    103   & 0.00  & 0.08  & 0.01  & 0.00  & 0.00  & 0.02 \\
    104   & 0.00  & 0.25  & 0.01  & 0.00  & 0.01  & 0.25 \\
    105   & 0.00  & 0.00  & 0.25  & 0.00  & 0.25  & 0.01 \\
    106   & 0.10  & 0.00  & 0.00  & 0.01  & 0.00  & 0.00 \\
    107   & 0.05  & 0.00  & 0.01  & 0.08  & 0.01  & 0.01 \\
    108   & 0.09  & 0.01  & 0.00  & 0.01  & 0.00  & 0.01 \\
    109   & 0.01  & 0.00  & 0.01  & 0.13  & 0.01  & 0.00 \\
    110   & 0.20  & 0.02  & 0.01  & 0.01  & 0.00  & 0.02 \\
 \hline
\end{longtable}